\documentclass[fleqn,usenatbib,useAMS]{mnras}
\usepackage{multicol}        % Multi-column entries in tables
\usepackage{bm}		% Bold maths symbols, including upright Greek
\usepackage{pdflscape}	% Landscape pages
\usepackage[T1]{fontenc}
\usepackage{ae,aecompl}
\usepackage{multirow}
\usepackage{siunitx}
\usepackage{textgreek}
\usepackage[dvipsnames]{xcolor}

\usepackage{float}
\usepackage{tikz}
\usepackage{listings}
\usepackage{algpseudocode}
\usepackage{algorithm}
\usepackage{lscape}
\usepackage{longtable,booktabs,tabu}

\DeclareRobustCommand{\VAN}[3]{#2}
\let\VANthebibliography\thebibliography
\def\thebibliography{\DeclareRobustCommand{\VAN}[3]{##3}\VANthebibliography}
\usepackage{enumitem}
\usepackage{academicons}
\usepackage{orcidlink}
\usepackage{xcolor,graphicx,xspace,color,longtable,tabu}
\usepackage{amssymb,amsmath,shadow,bezier,curves,rotating}
\usepackage{graphicx,chngcntr}
\usepackage{caption}
\usepackage{subcaption}
\graphicspath{ {./images/}}

\definecolor{orcidlogocol}{HTML}{A6CE39}
\usepackage[toc,titletoc]{appendix}
\usepackage {appendix}
\usepackage[flushleft]{threeparttable}

\newcommand{\mincir}{\raise
-3.truept\hbox{\rlap{\hbox{$\sim$}}\raise4.truept\hbox{$<$}\ }}
\newcommand{\magcisr}{\raise
-3.truept\hbox{\rlap{\hbox{$\sim$}}\raise4.truept\hbox{$>$}\ }}
\newcommand{\minmag}{\raise
-3.truept\hbox{\rlap{\hbox{$<$}}\raise5.truept\hbox{$<$}\ }}
\newcommand{\be}{\begin{equation}}
\newcommand{\ee}{\end{equation}}

\newcommand{\w}{\omega}

\newcommand{\ba}{\begin{eqnarray}}
\newcommand{\ea}{\end{eqnarray}}
\newcommand{\brr}{\begin{array}}
\newcommand{\err}{\end{array}}
\newcommand{\bc}{\begin{center}}
\newcommand{\ec}{\end{center}}

\title[Cluster Mass Function and $\sigma_8$-tension]{The Cluster Mass Function and the $\sigma_8$-tension}

\author[Alexandros Papageorgiou,  Manolis Plionis, Spyros Basilakos,  M. H. Abdullah] {A. Papageorgiou$^{1}$,  M. Plionis$^{1,2,3}$, S. Basilakos$^{3,4,5}$, M. H. Abdullah$^{6,7}$\\
\vspace{0.1cm} $^{1}$ Physics Department, University of Thessaloniki, Thessaloniki 54124,Greece \\
$^{2}$ National Observatory of Athens, Lofos Nymfon, 11851 Athens, Greece \\
$^{3}$ School of Sciences, European University Cyprus, Diogenes Street, Engomi, 1516 Nicosia, Cyprus \\
$^{4}$ Institute of Astronomy \& Astrophysics, National Observatory of Athens, Palaia Penteli 15230,
Athens, Greece \\
$^{5}$ Academy of Athens, Research Center for Astronomy and Applied Mathematics,
Soranou Efessiou 4, 115 27 Athens, Greece.\\
$^{6}$ Institute of Management and Information Technologies, Chiba University, 1-33, Yayoi-cho, Inage-ku, Chiba, 263-8522, Japan\\
$^{7}$ Astronomy Department, National Research Institute of Astronomy and Geophysics, Helwan, 11421, Egypt}

\begin{document}
  
\maketitle

\begin{abstract}

We use a large set of halo mass function (HMF) models in order to investigate their ability to represent the observational Cluster Mass Function (CMF), derived from the $\mathtt{GalWCat19}$ cluster catalogue, within the $\Lambda$CDM cosmology. We apply the $\chi^2$ minimization procedure to constrain the free parameters of the models, namely $\Omega_m$ and $\sigma_8$. We find that all HMF models fit well the observational CMF, while the Bocquet et. al. model provides the best fit, with the lowest $\chi^2$ value. Utilizing the {\em Index of Inconsistency} (IOI) measure, we further test the possible inconsistency of the models with respect to a variety of {\em Planck 2018} $\Lambda$CDM cosmologies, resulting from the combination of different probes (CMB - BAO or CMB - DES). We find that the HMF models that fitted well the observed CMF provide consistent cosmological parameters with those of the {\em Planck} CMB analysis, except for the Press $\&$ Schechter, Yahagi et. al., and Despali et. al. models which return large IOI values. The inverse $\chi_{\rm min}^2$-weighted average values of $\Omega_m$ and $\sigma_8$, over all 23 theoretical HMF models are: ${\bar \Omega_{m,0}}=0.313\pm 0.022$ and ${\bar \sigma_8}=0.798\pm0.040$, which are clearly consistent with the results of {\em Planck}-CMB, providing $S_8=\sigma_8\left(\Omega_m/0.3\right)^{1/2}= 0.815\pm 0.05$. Within the $\Lambda$CDM paradigm and independently of the selected HMF model in the analysis, we find that the current CMF shows no $\sigma_8$-tension with the corresponding {\em Planck}-CMB results.

\vspace {0.3cm}


\end{abstract}
\vspace {0.3cm}

\section{Introduction} \label{Intro}

The current cosmic structure formation paradigm assumes that dark matter (DM) halos form via gravitational instabilities of a primordial density field. Halos follow a roughly hierarchical merging process where halos with small masses form first and then more massive halos form through merging (e.g., \citealp{Springel05,Lacey93,Lacey94,Tormen04}). Galaxies and galaxy clusters, residing within these DM halos, arise from high peaks of an underlying initially Gaussian density fluctuation field \citep{Kaiser84,White91,Bardeen86}.
Although the details of the formation of galaxies within DM halos and the various feedback mechanisms are not yet fully understood, one can obtain insight into structure formation by studying the properties of DM halos and their evolution. The Halo Mass Function (HMF\footnote{Throughout the paper we use CMF for the cluster mass function derived from observations and HMF for the halo mass function computed by theoretical models.}), defined as the abundance of halos as a function of mass and redshift (e.g.\citealp{Press74,Sheth01,Vogelsberger14}), is one of these properties that play an important role in observational cosmology. The HMF is particularly sensitive to the matter density of the universe, $\Omega_m$ and the root mean square mass fluctuation at a scale of $8 h^{-1}$ Mpc at $z=0$ (e.g.,\citealp{Wang98}), as well as the evolution of bound structures from an early epoch to the late universe \citep{Press74,Bond91,Knox06}). 

One of the biggest challenges in utilizing the CMF as a cosmological probe is the difficulty of measuring cluster masses accurately. A variety of methods have been used such as, the virial mass estimator (e.g., \citealp{Binney87}) under the assumption of virial equilibrium, weak gravitational lensing (e.g., \citealp{Wilson96,Holhjem09}), or the application of Euler's equation on X-ray images of galaxy clusters (e.g., \citealp{Sarazin88}), under the assumption of hydrostatic equilibrium and spherical symmetry. However, these methods are observationally rather expensive since they require high-quality data and are also uncertain owing to the necessary assumptions made. Cluster masses can also be indirectly inferred from other observables, the so called mass proxies, which scale rather tightly with cluster mass. Among these mass proxies are X-ray luminosity, temperature, the product of X-ray temperature and gas mass (e.g., \citealp{Pratt09,Vikhlinin09,Mantz16a}), optical luminosity (e.g., \citealp{Yee03}), the velocity dispersion of member galaxies (e.g., \citealp{Biviano06,Bocquet15}), and cluster richness (e.g., \citealp{Pereira18,Abbott20,Abdullah23}). 
However, mass proxies are also hampered by uncertainties and scatter which may also introduce systematics in the estimation of cluster masses.  Furthermore, understanding the relationship between mass proxies and true mass is fundamental and using weak lensing mass calibration can aid to this effort (e.g., \citealp{Bocquet19}).

On the other hand, the determination of the theoretical HMF is also quite challenging. There are many approaches to calculate the HMF either analytically or via N-body simulations. The analytical predictions of the HMF are hampered by the difficulties of the non-linear gravitational collapse
which have not been fully addressed, except in the case of simple symmetries, while N-body simulations, in addition to being time consuming, do not provide a detailed understanding of the physical aspects of the mass accretion procedure.

\citet{Press74} (P-S) provided the first model of a linear analytic form of the HMF, under the assumption of spherical collapse and a primordial Gaussian density field in which structures grow via small perturbations. Early numerical simulations showed good agreements with the P-S formalism (cf. \citealp{Efstathiou79}). However, the advent of significant computer power and novel numerical techniques revealed deviations from the P-S formalism with overestimating/underestimating the number of halos at the high/low mass range (e.g. \citealp{Efstathiou88,Gross98,Governato99,Jenkins01,White02}), even if merging processes are included in the calculations (e.g.\citealp{Bond91,Lacey93}). Since then many authors attempted to provide an accurate HMF by extending the P-S formalism or by using N-body simulations. We briefly present some of these HMFs below. \citet{Sheth99} and \citet{Sheth01} (S-T) used an ellipsoidal collapse model to overcome the inaccuracies of P-S HMF. \citet{Jenkins01} found discrepancies with the S-T model and proposed an analytic fitting formula for the HMF. Their N-body simulations showed that the HMF was in agreement with the S-T analytic formula up to redshift $z=5$ in the mass range from galaxies to clusters. \citet{Tinker08} provided a HMF formula that improved previous approximations by $10-20\%$ using a large set of collisionless cosmological simulations. \citet{Crocce10} also proposed another HMF formula with an accuracy of $2\%$ up to redshift $z = 1$. 
\citet{Klypin11} used the {\em Bolshoi} simulation and found that although there was an agreement with the S-T model at low redshifts there were discrepancies at high redshifts.
Also \citet{Bhattacharya11} provided a HMF that was accurate to $2\%$ at redshift $z = 0$ and $10\%$ at redshift $z = 2$. \citet{Angulo12} used the Millenium-XXL simulation to obtain a HMF which was accurate to better than $5\%$ over the mass range used in this study. \citet{Comparat17,Comparat19} used the Multidark simulation to revisit the HMF which was accurate at $<2\%$ level. \citet{Bocquet16} used hydrodynamical and dark matter only simulations in order to calibrate the HMF. Additionally, \citet{Bocquet20} constructed an emulator for estimating the HMF, which was better than $<2\%$ for DM halo masses of $10^{13} - 10^{14}h^{-1}M_{\odot}$ and $<10\%$ for $10^{15}h^{-1}M_{\odot}$. Recently, \citet{Shirasaki21} used high-resolution N-body simulations and provided a new fitting formula for HMF. Their model appears accurate at a $5\%$ level, except for the large masses at $z \leq 7$. They showed that the S-T model overestimates the halo abundance at $z=6$ by $20-30\%$.

An important claim regarding the HMF is its apparent universality in the sense that the same HMF parameters fit different redshifts and cosmologies (e.g. \citealp{Jenkins01,Sheth99,Reed03,Warren06,Watson13}). However, various authors have rejected this claim (e.g. \citealp{White02,Reed07,Crocce10,Bhattacharya11,Courtin11, Tinker08}). Recently, \citet{Diemer20}, studying the universality of the HMF, seem to confirm that for the $\Lambda$CDM and for a wide range of halo mass definitions there is a varying level of non-universality that increases with peak height and therefore also with cluster mass.
However, they also found that splashback radius-based HMF is universal to $10\%$ at $z\leq2$ in agreement with some previous studies that adopted more traditional definitions of the halo radius (e.g. \citealp{Warren06})\footnote{The splashback halo radius is defined as the halo boundary where particles reach the apocenter of their first orbit; \citet{Diemer14}.}

Thus, it appears that the universality or non-universality of the HMF parameters depends on how halos are defined. Halos identified using the Friend-of-Friend algorithm (FoF) provide almost a universal HMF, while halos identified using the spherical overdensity (SO) approach returns non-universal HMF, especially at high redshifts. These discrepancies might be caused by the fact that the FOF algorithm links objects before they merge.

Although the $\Lambda$CDM model fits a wide variety of cosmological data with high accuracy, tensions have emerged between the values of the Hubble constant, $H_0$, and the  normalization of the matter power-spectrum, $\sigma_8$ (ie., the root mean square amplitude of the matter fluctuations in spheres of 8$h^{-1}$ Mpc) derived, within the $\Lambda$CDM paradigm, using the Planck CMB temperature fluctuation data (corresponding to redshifts $z\sim 1100$)) and low-redshift ($z\lesssim 0.15$) data (e.g. Cepheid-based distance indicator, cosmic shear, clusters of galaxies etc).

The $\sigma_8$-tension is equivalent to the $S_8$ tension, where $S_8=\sigma_8 (\Omega_m/0.3)^{1/2}$ is the so-called Cluster Normalization condition.
\citet{Aghanim20} estimated the amplitude of the cosmological perturbations from the CMB as $\sigma_8 = 0.811\pm0.006$, with corresponding $S_8=0.834\pm 0.016$, while a wide range of data based on weak lensing and galaxy clustering at lower redshifts  provide systematically lower values of $S_8$. For example, \citet{Schuecker03} found $\sigma_8 = 0.711\pm^{0.039}_{0.031}$ with inferred $S_8\simeq 0.76\pm 0.05$ (using the REFLEX X-ray cluster data), while the most recent analysis of X-ray clusters, that of the {\sc eROSITA} Final Equatorial Depth Survey (eFEDS) with uniform weak-lensing cluster mass constraints (from the Hyper Suprime-Cam survey)  provided  $S_8\simeq 0.753\pm0.06$  (\citealp{2023MNRAS.522.1601C}).
\citet{Heymans21}, based on the Kilo-Degree Survey (KiDS-1000), found $S_8=0.766^{+0.020}_{-0.014}$, a tension of $\sim 3\sigma$ with the results of the CMB analysis, while the recent joint cosmic shear analysis of the Dark Energy Survey (DES Y3) and the KiDS-1000  resulted in $S_8=0.790^{+0.018}_{-0.014}$  (\citealp{2023OJAp....6E..36A}). Similarly, \citet{Bocquet19} analysing SPT-SZ cluster data with weak-lensing mass calibration and a procedure where cosmology, scaling relations and other nuisance parameters were fit simultaneously, found  $S_8=0.745 \pm 0.042$.
The recent {\em Planck} SZ cluster analysis of \citet{2020A&A...641A...1P} provided $S_8\simeq 0.78^{+0.01}_{-0.03}$ (for a mass bias $1-b=0.8$), which however increases to $\simeq 0.86$ for a lower mass bias, while \citet{Zubeldia19} (using also {\em Planck} SZ - MMF3 cluster sample) found $\sigma_8=0.76\pm0.04$ with inferred $S_8\simeq 0.797\pm 0.04$.

The tension in $S_8$, together with that of the Hubble constant, $H_0$, could be an indication for the necessity of a new cosmological paradigm, beyond the canonical $\Lambda$CDM. However, as far as the $S_8$ tension is concerned, it is evident from the above that it shows up in some datasets and not in others, which hints towards unknown systematic errors entering in some  analyses procedures. Amon \& Efstathiou (2022) argue for a possible consistency of the weak-lensing analyses results with those of {\em Planck} CMB if the matter power spectrum is suppressed more strongly on non-linear scales.

In the present paper we compare a large set of 23 HMF models with the observationally determined CMF from \citet{Abdullah20b}, in order to explore the validity of these HMF models, within the $\Lambda$CDM cosmological background, and also to constrain the two free cosmological parameters of the models, namely, the matter density parameter, $\Omega_m$ and the $\sigma_8$ normalization factor. Recently, a different procedure was followed by \citet{Driver22}, who reconstructed the HMF through a model-free empirical approach using the GAMA galaxy group data, finding a good agreement with the expectations of the $\Lambda$CDM model.

The outline of this paper is as follows. 
In Section \ref{sec:CMF} we describe the CMF obtained from \citet{Abdullah20b}.  In Section \ref{sec:HMF} we introduce the main elements of the theoretical HMF models. In Section \ref{sec:Res} we present the results of our statistical analysis regarding the cosmological parameters $\Omega_{m,0}$ and $\sigma_8$, while in Section \ref{sec:Conc} we draw our conclusions. Throughout this work we use, when necessary, the value of $h=H_0/100=0.678$. We have, however, verified that our results are insensitive to the exact value of $h$ within the current range defined by the so-called Hubble constant tension.

\section{Cluster of Galaxies Mass Function - CMF}
\label{sec:CMF}
There have been various attempts to derive the observational CMF either using the 2PIGG galaxy groups of the 2 degree Field Galaxy Redshift Survey by \citet{Eke2006}, the SDSS-DR10 galaxy groups by \citet{Tempel2014}, the ROSAT-based REFLEX-II X-ray cluster catalogue by \citet{Bohringer17}, or recently the GAMA galaxy group data by \citet{Driver22}.

In our current work we use the recent SDSS-DR13 \citep{Albareti17} $\mathtt{GalWCat19}$ cluster catalogue \citep{Abdullah20a}. Using photometric and spectroscopic databases from SDSS-DR13, \citet{Abdullah20a} extracted data for 704,200 galaxies.
These galaxies satisfied the following set of criteria: spectroscopic detection, photometric and spectroscopic classification as a galaxy (by the automatic pipeline), spectroscopic redshift between 0.01 and 0.2 (with a redshift completeness >0.7; \citealp{Yang07,Tempel14}), $r$-band magnitude
(reddening-corrected) <18, and the flag {\em SpecObj.zWarning} having a value of zero, indicating a well-measured redshift. \citealp{Abdullah20a} identified galaxy clusters utilizing the Finger-of-God effect.
Then, they assigned cluster membership using the {\em GalWeight} technique \citep{Abdullah18}. The {\em GalWeight} technique has been tested with N-body simulations ({\em MDPL2} and {\em Bolshoi}) and it has been shown that it is $98\%$ accurate in assigning cluster membership. Additionally, the masses of each cluster were calculated by applying the virial theorem (e.g. \citealp{Binney87,Abdullah11}) and correcting for the surface pressure term \citep{The86}. The virial theorem avoids any assumption about the internal physical processes associated with baryons (gas and galaxies).
More specifically, the cluster mass was calculated at the virial radius at which the density equals $\rho = \Delta_{200}\rho_c$, where $\rho_c$ is the critical density of the universe and $\Delta_{200}=200$. The uncertainty of the virial mass is calculated using the limiting fractional uncertainty $=(1/\pi)\sqrt{2\ln{N}/N}$ \citep{1981ApJ...244..805B}. The systematic uncertainties in the computation of mass are due to the: (i) assumption of virial equilibrium, projection effects, velocity anisotropies in galaxy orbits, etc, (ii) presence of substructure and/or nearby structures, (iii) presence of interlopers in the cluster frame, (iv) uncertainties in the identification of the cluster center.

The $\mathtt{GalWCat19}$ cluster catalogue contains 1800 clusters in the redshift range of $0.01 \leq z \leq0.2$ and in a mass range of $4\times 10^{13}\leq M/ h^{-1} M_{\odot}\leq2\times 10^{15}$.

The $\mathtt{GalWCat19}$ cluster catalogue is incomplete in redshift at $z > 0.085$ (or comoving distance $D > 265 h^{-1}$) and to correct for this incompleteness, each cluster with a comoving distance $D > 265 h^{-1}$ Mpc is weighted by the inverse of the selection function $S(D)$, estimated in \citet{Abdullah20b} by using the normalized cluster number density and fitting it to an exponential function. The final selection function used is given by:
\begin{equation}\label{eq:wp1}
S(D) = a \times \exp\left[- \left(\frac{D}{b}\right)^{\gamma}\right] \;,
\end{equation} where $a = 1.1\pm0.12$, $b = 293.4\pm 20.7 h^{-1}$ Mpc and $\gamma = 2.97 \pm 0.9$. 
Systematic effects due to the uncertainties in the selection function were investigated by  \citet{Abdullah20b} to conclude that they leave mostly unaffected the main results of the analysis. Furthermore, they studied the effect of the inverse selection function weighting at larger redshifts, where the cluster sampling is low and the shot-noise effects large, an analysis which resulted in identifying the optimum redshift-range within which the possible overcorrection is minimized. To this end \citet{Abdullah20b} restricted the sample to a maximum comoving distance of $D \leq 365 h^{-1}$ Mpc, ie., $z\leq 0.125$. As a further test \citet{Abdullah20a} compared the CMF derived from the $\mathtt{GalWCat19}$ (after weighting to recover volume completeness) with the HMF derived from the {\em MDPL2} simulation. The comparison indicated that the sample is complete in mass for $\log(M)\geq 13.9 h^{-1} M_{\odot}$. Additionally, as shown in Figure 1 of \citet{Abdullah20b}, this mass threshold is roughly the same for a variety of cosmologies. The final cluster subsample (SelFMC, presented by \citealp{Abdullah20b}) contains 843 galaxy clusters in the redshift range $0.01<z<0.125$ with $\log{M} \geq 13.9 h^{-1} M_{\odot}$.

The CMF mass function is estimated using: 
\begin{equation}\label{eq:wp2}
\frac{{\rm d}n(M)}{{\rm d}\log{M}} = \frac{1}{{\rm d}\log{M}}\sum_{i}\frac{1}{V}\frac{1}{S(D_{i})} \;,
\end{equation}
where $D_{i}$ is the comoving distance and $V$ is the comoving volume given by:
\begin{equation}\label{eq:wp3}
V = \frac{4\pi}{3}\frac{\Omega_{\rm survey}}{\Omega_{\rm sky}}(D^{3}_{2} - D^{3}_{1}) \;,
\end{equation}
where $\Omega_{\rm sky} = 41.253$ deg$^{2}$, $\Omega_{\rm survey} = 11.000$ deg$^{2}$ and
$D_{1}$, $D_{2}$ are the minimum and maximum comoving distances of the sample.

It is worth mentioning that the $\mathtt{GalWCat19}$ catalogue is one of the largest available
spectroscopic cluster samples for which the determination of the cluster center, cluster redshift and membership assignment are of high accuracy. In Table \ref{tab01} we present the precise numerical values of the observationally determined SelFMC mass function data, with the corresponding errors, that are used in our current analysis. The width of each logarithmic mass bin is equal to 0.155.

\begin {table}
\begin{center}
    \begin{tabular}{ c c    }
    \hline 
    \\[0.1ex]
$\log$M &    $\log$(dn/d$\log$M)                                 \\[1.ex]  \hline 
$13.9775$   & $-4.165^{+0.026}_{-0.026}$  \\     
$14.1325$   & $-4.179^{+0.027}_{-0.030}$      \\
$14.2875$   & $-4.382^{+0.034}_{-0.037}$     \\ 
$14.4425$   & $-4.594^{+0.053}_{-0.060}$     \\ 
$14.5975$   & $-5.093^{+0.074}_{-0.089}$     \\ 
$14.7525$   & $-5.501^{+0.132}_{-0.189}$     \\ 
$14.9075$   & $-5.753^{+0.148}_{-0.229}$     \\ 
$15.0625$   & $-6.892^{+0.232}_{-0.537}$     \\  \hline 
    \end{tabular}
\end{center}
\caption {The measured Cluster Mass Function data of the SelFMC subsample from \citet{Abdullah20b}. Note that the width of each logarithmic mass bin is equal to 0.155.}
\label{tab01}
\end{table}

\section{Halo Mass Function Models}
\label{sec:HMF}
In this section we present a large number of the mass function models found in the literature.
The number of dark matter halos per comoving volume per unit mass is given by:
\begin{equation}\label{eq:wp4}
\frac{{\rm d}n(M)}{{\rm d}\ln{M}} = f(\sigma)\frac{\rho_0}{M} \left|\frac{{\rm d}\ln\sigma}{{\rm d}\ln M}\right| \;,
\end{equation}
where $\sigma^2(M,z)$ is the mass variance of the smoothed linear density field, given in 
Fourier space by:
\begin{equation}\label{eq:wp5}
\sigma(M_h,z)=\left[\frac{D^2(z)}{2\pi^2}\int_0^{\infty} k^2P(k) W^2(kR)dk\right]^{1/2} \;,
\end{equation}
where $D(z)$ is the growth factor of matter fluctuations in the linear regime,
$W(kR) = 3[\sin(kR) - kR\cos(kR)]/(kR)^3$ is 
the Fourier transform of the top-hat smoothing kernel of radius R. The radius is given by
$R = [3M_h/(4\pi\rho_{m})]^{1/3}$ with $M_h$ the mass of the halo 
and $\rho_{m}$ the mean matter density of the universe at the present time.
The quantity $P(k, z)$ is the CDM linear power spectrum given by $P(k, z)=P_0k^nT^2(k)D^2(z)$ where 
$n$ is the spectral index of the primordial power spectrum and $T(k)$ is the CDM transfer function 
provided by \citet{Eisenstein98}:
\begin{equation}\label{eq:wp6}
T(k) = \frac{L_0}{L_0 +C_0q^2} \;,
\end{equation}
with $L_0={\rm ln}(2e+1.8q)$, $e=2.718$, $C_0 = 14.2+\frac{731}{1+62.5q}$ 
and $q =k/\Gamma$, with $\Gamma$ being the shape parameter given by \citet{Sugiyama95}:
$$
\Gamma= \Omega_{m}h{\rm exp}(-\Omega_{b}-\sqrt{2h}\frac{\Omega_{b}}{\Omega_{m}}).
$$

The normalization of the power spectrum is given by:
\begin{equation}\label{eq:wp7}
P_0 = 2\pi^2\sigma_8^2\left[\int_0^{\infty}T^2(k)k^{n+2}W^2(kR_8)dk\right]^{-1} \;,
\end{equation}
where $\sigma_{8}\equiv \sigma(R_{8},0)$ is the present value of the mass variance at $8 h^{-1}$ Mpc.
In the original P-S formalism the term $f(\sigma)$, which appears in eq.(4), is given by:
$ f_{P-S}(\sigma)=\sqrt{2/\pi} (\delta_c/\sigma) \exp{(-\delta_c^2/2\sigma^2)}$ with $\delta_c$ the linearly extrapolated density threshold above which structures collapse. For the HMF models considered here, the form of $f(\sigma)$ and its parameters are listed in Table \ref{tab:fittingfunctions}. 

\begin{table*}
\centering
  \caption[]{Compilation of Fitting Functions}
  \label{tab:fittingfunctions}	
    \scriptsize
    \begin{tabular}{ll}
    \hline
    \textsc{Ref.} & \textsc{Fitting Function $f(\sigma)$}\\ 
    \hline
    
    \citet{Press74} & $f_{\rm PS}(\sigma) = \sqrt{\frac{2}{\pi}}\frac{\delta_c}{\sigma}\exp\left[-\frac{\delta_c^2}{2\sigma^2}\right]$\\
      
    \citet{Sheth01} & $f_{\rm ST}(\sigma) =A\sqrt{\frac{2a}{\pi}}\left[1+\left(\frac{\sigma^2}{a\delta_c^2}\right)^p\right]\frac{\delta_c}{\sigma}\exp\left[-\frac{a\delta_c^2}{2\sigma^2}\right]$, $A=0.3222$, $a=0.75$, $p=0.3$\\
      
    \citet{Jenkins01} & $f_{\rm J}(\sigma) = 0.315 \exp \left[|\ln \sigma^{-1}+0.61|^{3.8}\right]$\\
        
    \citealp{White02} &
    $f_{\rm Wh_{FOF}}(\sigma) = f_{\rm ST}(\sigma)$,
    $\alpha = 0.70$, $p = 0.29$ \\
  
    \citet{Reed03} & $f_{\rm R03}(\sigma) =f_{ST}(\sigma) \exp\left[\frac{-0.7}{\sigma \cosh(2\sigma)^5}\right]$\\

    \citet{Yahagi_2004} &
    $f_{\rm Y}(\sigma)=f_{\rm ST}(\sigma)$, $A =  0.301$, $\alpha = 0.664$, $p = 0.321$\\

    \citet{Warren06} & $f_{\rm W}(\sigma) =0.7234 \left(\sigma^{-1.625}+0.2538\right) \exp\left[\frac{-1.1982}{\sigma^2}\right]$\\
  
    \citet{Reed07} & 
    $f_{\rm R07b}(\sigma) = A\sqrt{2\alpha}{\pi}\left[1 + \left(\frac{\sigma^2}{\alpha\delta_c^2}\right)^p + 0.2G_1\right]\frac{\delta_c}{\sigma}\exp\left[-\frac{c\alpha\delta_c^2}{2\sigma^2}\right]$, $A = 0.3222$, $\alpha = 0.707$, $p = 0.3$, $c =1.08$, \\&
    $G_1 = \exp\left[-\frac{(\ln\sigma^{-1} - 0.4)^2}{2(0.6)^2}\right]$\\
    
    \citet{Reed07} & 
    $f_{\rm R07a}(\sigma)= A \sqrt{\frac{2\alpha}{\pi}}[1 + \left(\frac{\sigma^{2}}{\alpha\delta_c^2}\right)^p + 0.6G_1 + 0.4G_2] \times 
    \frac{\delta_c}{\sigma}\exp\left[-\frac{c\alpha\delta_c^2}{2\sigma^2} - \frac{0.03}{(n_{\rm eff} + 3)^2}\left(\frac{\delta_c}{\sigma}\right)^{0.6}\right]$,\\ 
    &$A = 0.3222$, $\alpha = 0.707$, $p = 0.3$, $c =1.08$,\\
    &$n_{\mathrm{eff}} = 6\frac{d \log \sigma^{-1}}{d \log M} -3$, 
    $G_1(\sigma) =  \exp\left[-\frac{(\ln\sigma^{-1} - 0.4)^2}{2(0.6)^2}\right]$,
    $G_2(\sigma) = \exp\left[-\frac{(\ln\sigma^{-1} - 0.75)^2}{2(0.2)^2}\right]$\\
      
    \citet{Tinker08} & 
    $f_{\rm T}(\sigma,z) =A(z)\left[\left(\frac{b(z)}{\sigma}\right)^{a(z)} + 1\right] \exp\left[-\frac{c}{\sigma^2}\right]$, \\&
    $A(z) = 0.186\left(1+z\right)^{-0.14}$,	$a(z) = 1.47\left(1+z\right)^{-0.06}$, $b(z) = 2.57\left(1+z\right)^{-\alpha}$, $c = 1.19$,
    $\alpha = \exp\left[-\left(\frac{0.75}{\ln(\Delta_{\mathrm{vir}}/75)}\right)^{1.2}\right]$\\

    \citet{Manera_2009} &       
    $f_{\rm M}(\sigma)=f_{\rm ST}(\sigma)$,
    $A = \left[1 + 2^{-\alpha}\Gamma(1/2 - \alpha)/\Gamma(1/2)\right]^{-1}$, $\alpha = 0.687$, $p = 0.261$.\\

    \citet{bagla09} &
    $f_{\rm Ba}(\sigma)=f_{\rm ST}(\sigma)$
    $\alpha = 0.677$, $p = 0.25$, $A = [1 + 2^{-\alpha}\Gamma(1/2 - \alpha)/\sqrt{\pi}$\\
    
    \citet{Maggiore_2010} &
    $f_{\rm M}(\sigma) = (1 - \widetilde{\kappa})\left(\frac{2\alpha}{\pi}\right)^{1/2}\frac{\delta_c}{\sigma}\exp\left[-\frac{\alpha\delta_c^2}{2\sigma^2}\right] +\widetilde{\kappa}\frac{\delta_c\sqrt{\alpha}}{\sigma\sqrt{2\pi}}\Gamma\left(0, \frac{\alpha\delta_c^2}{2\sigma^2}\right)$,
    $\widetilde{\kappa}=\frac{\kappa}{D_B}$, $\kappa = 0.4592-0.0031R$, $D_B=0.25$, $\alpha=0.8$
    \\
      
    \citet{Crocce10} & 
    $f_{\rm Cr}(\sigma,z) =A(z)\left(\sigma^{-a(z)} + b(z)\right) \exp\left[-\frac{c(z)}{\sigma^2}\right]$, $A(z) = 0.58\left(1+z\right)^{-0.13}$,	\\&
    $a(z) = 1.37\left(1+z\right)^{-0.15}$, $	b(z) = 0.3\left(1+z\right)^{-0.084}$, $c(z) = 1.036\left(1+z\right)^{-0.024}$\\
      
    \citet{Courtin10} & 
    $f_{\rm Co}(\sigma) =f_{\rm ST}(\sigma)$, 
    $A = 0.348$,	$a = 0.695$, $p = 0.1$\\
      
    \citet{Bhattacharya11} & 
    $f_{\rm B}(\sigma,z) = A\sqrt{\frac{2}{\pi}}\exp\left[-\frac{a\delta_c^2}{2\sigma^2}\right]\left[1+\left(a\frac{\delta_c^2}{\sigma^2}\right)^{-p}\right]\left(\frac{\delta_c^2}{\sigma^2}\sqrt{a}\right)^q$, 
    $A = 0.333\left(1+z\right)^{-0.11}$,	\\
    &$a = 0.788\left(1+z\right)^{-0.01}$, $	p = 0.807$, $q =1.795$\\
      
    \citet{Angulo12} & 
    $f_{\rm A}(\sigma) =A\left[\left(\frac{b}{\sigma}\right)^a+1\right] \exp\left[-\frac{c}{\sigma^2}\right]$, $\left(A,a,b,c\right) = \left(0.201,1.7,2.08,1.172\right)$ \\
      
    \citet{Watson13} & 
    $f_{\rm Wa_{FOF}}(\sigma) = f_{\rm T}(\sigma)$,
    $A = 0.282$,	$a = 1.406$, $	b = 2.163$, $c = 1.21$\\

    \citet{Despali_2015} &
    $f_{\rm D}(\sigma)=f_{\rm ST}(\sigma)$
    $A = 0.287$, $\alpha = 0.903$, $p = 0.322$\\

    \citet{Bocquet16} &
    $f_{\rm Bo}(\sigma,z) =f_{\rm T}(\sigma,z)$, 
    $A(z) = 0.228(1+z)^{0.285}$,
    $\alpha(z) = 2.15(1+z)^{-0.058}$, $b(z) = 1.69(1+z)^{-0.336}$,
    $c(z) = 1.30(1+z)^{-0.045}$ \\

    \citet{Benson_2017} &
    $f_{\rm Be}(\sigma)=f_{\rm ST}(\sigma)$, $A = 0.3310$, $p = 0.1869$, $\alpha = 0.7722$\\
 
    \citet{Comparat_2017} &
    $f_{\rm C}(\sigma)=f_{\rm B}(\sigma)$,
    $A_0 = 0.3241$, $\alpha_0 = 0.897$, $p_0 = 0.624$, $q_0 = 1.589$\\
    
    \citet{Shirasaki21} &
    $f_{\rm S}(\sigma,z)=f_{\rm B}(\sigma,z)$, 
    $p(z) = 0.692$, $q(z) = 1.611(1+z)^{0.12}$,\\& $\alpha(z) = 0.94 (1+z)^{-0.02}\left(1.686/\delta_c\right)^{2} \left[1 - 0.015 (z/1.5)^{0.01}\right]^{-1}$, $A(z) = 0.325 - 0.017z$\\
          
    \hline
    \end{tabular}
\end{table*}

\begin{figure*}\hspace{0cm}
\centering
\includegraphics[width=\linewidth]{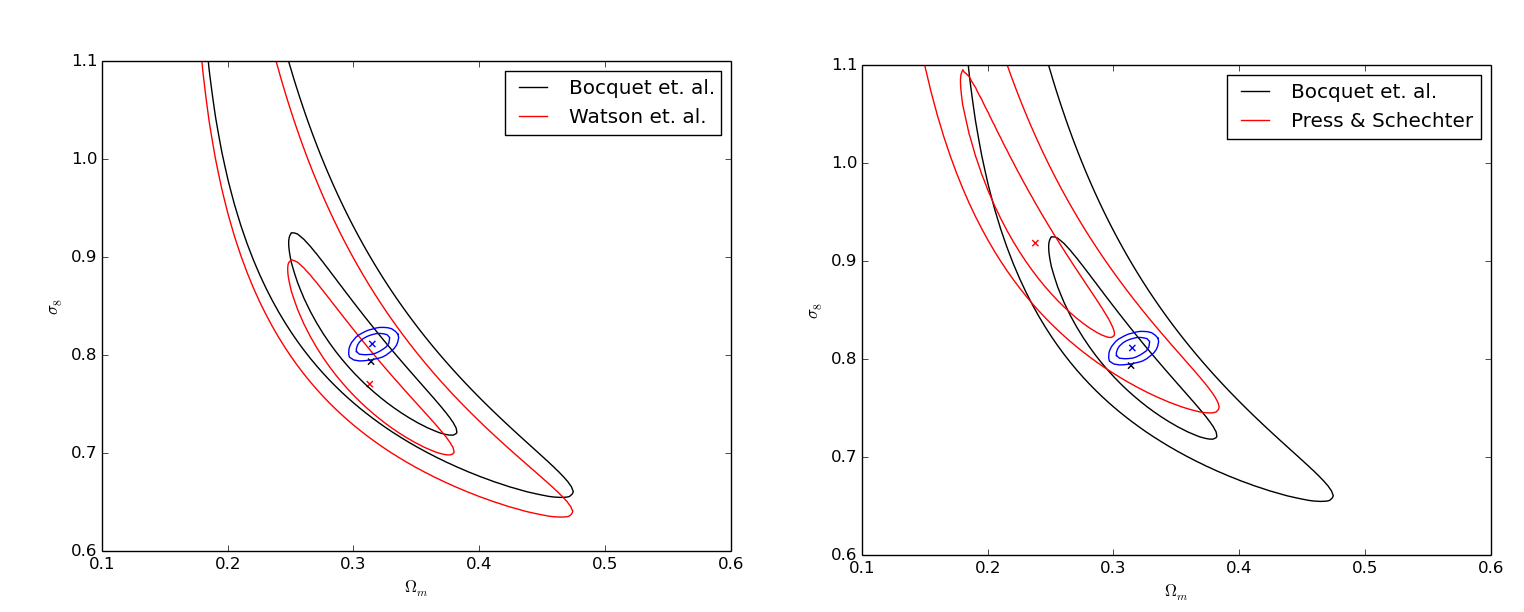}
\caption{Contour plot of the $1\sigma$ and $3\sigma$ confidence levels for the two fitted parameters, $\sigma_8$ and $\Omega_m$. The left plot presents the contours of \citet{Bocquet16} (black line) and the \citet{Watson13} (FOF) (red line) HMFs which have the lowest $\chi^2_{\rm min}$. The right panel presents the contour plots of \citet{Bocquet16} HMF (black line), which has the lowest $\chi^2_{\rm min}$, and \citet{Press74} HMF (red line), which has the highest $\chi^2_{\rm min}$. 
The blue contour presents the $1\sigma$ and $2\sigma$ confidence levels of the CMB analysis of the \citet{Aghanim20}.}
    \label{f2}
\end{figure*}

Since the majority of the HMF models appear to be either universal or the deviations from universality are rather small for the $\Lambda$CDM cosmology, for simplicity we use the exact parameter values from each corresponding paper.

Additionally, it is important to note that (a) the $\mathtt{GalWCat19}$ clusters are basically local, with a narrow redshift range and thus there is no significant redshift evolution (\citealp{Abdullah20b}), and (b) wherever we had the option we used HMF parameters relevant to the $M_{200}$ halo mass definition. However, the majority of the HMF models use an FOF halo mass definition (with $b=0.2$), corresponding to $\Delta = 178$, which indeed provides halo masses almost identical to $M_{200}$; see for example \cite{2001A&A...367...27W}.
Furthermore, using the subset of HMF models for which the values of halo mass were provided for both $\Delta_{\rm crit}$ and $\Delta_{\rm mean}$, we found that the differences of the HMF ranges between $0.5-1.5\%$ for low masses and $4.5-5.5\%$ for high masses, depending on the cosmological parameters. Repeating our cosmological analysis does not alter significantly our results (e.g. for the Tinker model we found $\delta\Omega_m \sim 0.3\%$ and $\delta\sigma_8 = 0.7\%$).

\section{Results}
\label{sec:Res}
\subsection{Fitting HMF Models to the observed CMF}
To test the validity and to quantify the free parameters of the aforementioned mass function models 
we perform a standard $\chi^2$-minimization procedure between the observational CMF derived from the SelFMC cluster subsample and the expected theoretical HMF in the specific mass range covered by the CMF (c.f. \citealp{Abdullah20b}; \citealp{2021MNRAS.502.3942H}). The $\chi^2$ function is defined as follows:

\begin{equation}
\chi^2 = \displaystyle\sum^{8}_{i=1}\frac{\left[\log\left(\frac{{\rm d}n}{{\rm d}\log M}_{\rm obs}\right)-
\left(\log\frac{{\rm d}n}{{\rm d}\log M}_{\rm th}(\textbf{p})\right) \right]^2}{\sigma_{i}^{2}+\sigma_{M}^{2}} \;,
\end{equation}
where the vector ${\bf p}$ contains the main  free parameters of the theoretical HMF, namely ${\bf p} = (\Omega_m, \sigma_8)$.
Note that $\sigma_{i}$ is the CMF uncertainty (see Table \ref{tab01}) and $\sigma_M=0.075$ corresponds to the half-width of the logarithmic cluster mass-bin.
 
Since we have non-symmetric uncertainties for the CMF we use a weighted uncertainty scheme in the $\chi^2$ function, following \citet{Barlow04}, according to which:
\begin{equation}
  %\sigma_i = \sigma_1 + \sigma_2\left[\log\left(\frac{{\rm d}n}{{\rm d}\log M}_{\rm obs}\right)-
\sigma_i = \sigma_1 + \sigma_2\left[\log\left({\rm d}n/{\rm d}\log M_{\rm obs}\right)-
\left(\log{\rm d}n/{\rm d}\log M_{\rm th}(\bf{p})\right)\right]
\end{equation}
where $\sigma_1 = 2\sigma_p \sigma_n/(\sigma_p + \sigma_n)$ and $\sigma_2 = (\sigma_p - \sigma_n)/(\sigma_p + \sigma_n)$.
Note that for $\sigma_p = \sigma_n$ we get the symmetric error weighting (see \citealp{Barlow04} for details).

Furthermore, we utilize the corrected {\em Akaike information criterion}
\citep{Akaike74,Sugiura78} which for the current application, having a small number of data points, it transforms to (see \citealp{Liddle07}):
\begin{equation}\label{AIC}
{\rm AIC}_{c}=\chi^{2}_{\rm min}+2k+\frac{2k(k+1)}{N-k-1} \;,
\end{equation}
with $N$ the number of data points and $k$ is the number of free parameters.
A small value of AIC indicates a better model-data fit.
Furthermore, we also examine the Akaike information criterion difference between a HMF model pair, namely
$\Delta$AIC=AIC$_{c,x}$-AIC$_{c,y}$, in order to evaluate the relative ability of any two HMF models to reproduce the data.
The larger the difference $|{\rm \Delta AIC}|$, the higher the evidence of inconsistency (with the comparison model) of the model with higher value of AIC$_{c}$. A difference between $4\le |{\rm \Delta AIC}| \le 7$ is a positive indication of the previous statement \citep{burnham2004},
$|{\rm \Delta AIC}| \ge 10$ indicates a strong such evidence, and
$|{\rm \Delta AIC}| \le 2$ is an indication that the two comparison models are rather consistent.

In Figure (\ref{f2}) we present our results of the $\chi^2$ minimization procedure. The figure shows the contours of the $1\sigma$ and $3\sigma$ confidence levels for the two fitted parameters, $\sigma_8$ and $\Omega_m$. The left panel presents the contours of the \citet{Bocquet16} (black line) and the \citet{Watson13} (FOF) (red line) HMF models, which have the lowest $\chi^2_{\rm min}$ values. The right panel presents the contours of the former model (black line) and of the \citet{Press74} model (red line), which has the highest $\chi^2_{\rm min}$ value.

In Table \ref{tab:growth2} we present the statistical results of the $\chi^2$-minimization procedure, with the first and second columns listing the derived $\Omega_m$ and $\sigma_8$ and their corresponding $1\sigma$ uncertainties. It is evident that the values of both parameters, for the majority of the HMF models, are consistent with the recent CMB-Planck results \citep{Aghanim20} (see Section \ref{sec:Conc} for further discussion).

In the following columns we present
the goodness of fit statistics, namely the $\chi_{\rm min}^{2}$ and the AIC$_{c}$. As it is evident by taking into account the resulting $\chi_{\rm min}^{2}$ values, all the HMF models fit the data at an acceptable level (with a reduced $\chi_{\rm min}^2\sim 1.3-1.5$), with the best model (lowest $\chi_{\rm min}^2$ value) being that of Bocquet et al (2016). Note that if the CMF uncertainties were larger by 15\% we would have obtained a reduced $\chi^2_{\rm min}\simeq 1$.
In the last column we present the difference $\Delta$AIC=AIC$_{c,{\rm Bocquet}}$-AIC$_{c,y}$ between the best HMF model (Bocquet et al) and any other model ($y$) and find for all HMF models that $\Delta$AIC=AIC$_{c,{\rm Bocquet}}$-AIC$_{c,y} < 2$, which means that all the models are consistent with the reference model. We conclude that with the current observational CMF data we cannot distinguish, at a statistically acceptable level, between the available HMF models.

\begin{table*}%[ht]
\caption[]{Results for the fit of the observed CMF to a large range of theoretical HMF: The $1^{st}$ column indicates the HMF model, the
  $2^{nd}$ and the $3^{rd}$ column provides the fitted $\Omega_m$ and $\sigma_8$ respectively. The remaining columns present the goodness-of-fit statistics $\chi^{2}_{\rm min}$, AIC$_{c}$ and $|\Delta$AIC$|=|{\rm AIC}_{c,\rm Bocquet}-{\rm
    AIC}_{c,y}|$ with the index $y$ corresponding to the indicated comparison model.}%, and the last column the FoM.}

\tabcolsep 4.5pt
\vspace{1mm}
\begin{tabular}{cccccc} \hline \hline
HMF Model & $\Omega_m$ & $\sigma_8$&  $\chi_{\rm min}^{2}$ & ${\rm AIC_{c}}$&       $|\Delta$AIC$|$  
 \vspace{0.05cm}\\ \hline
 Bocquet et. al.(2016)& 		$0.314^{+0.024}_{-0.021}$  &$0.793^{+0.037}_{-0.028}$& 8.103 & 14.503& -
 \vspace{0.15cm}\\ 
Watson et. al. (2013-FOF)& 		$0.313^{+0.024}_{-0.021}$   &$0.770^{+0.036}_{-0.028}$  & 8.226 & 14.626 & 0.123 
 \vspace{0.15cm}\\
Jenkins et. al. (2001)& 			$0.294^{+0.023}_{-0.020}$   &$0.800^{+0.037}_{-0.029}$& 8.250 & 14.650 & 0.147 
\vspace{0.15cm}\\  
Angulo et. al. (2012)& 				$0.297^{+0.023}_{-0.020}$   &$0.786^{+0.036}_{-0.028}$& 8.451 & 14.851 & 0.348
\vspace{0.15cm}\\
Bhattacharya et. al. (2011)& 	$0.310^{+0.024}_{-0.021}$   &$0.762^{+0.036}_{-0.027}$& 8.476 & 14.876 & 0.373 
 \vspace{0.15cm}\\
Warren et. al. (2006)& 				$0.297^{+0.023}_{-0.020}$   &$0.799^{+0.037}_{-0.029}$& 8.478 & 14.878 & 0.375
\vspace{0.15cm}\\
Shirasaki et. al. (2021)& 		$0.318^{+0.024}_{-0.021}$   &$0.857^{+0.041}_{-0.031}$& 8.562 & 14.962 &0.459
\vspace{0.15cm}\\
Comparat et. al. (2019)& 			$0.316^{+0.024}_{-0.021}$   &$0.830^{+0.039}_{-0.030}$& 8.576 & 14.976 &0.473
\vspace{0.15cm}\\
Tinker et. al.(2008)& 				$0.302^{+0.023}_{-0.021}$   &$0.809^{+0.038}_{-0.030}$& 8.604 & 15.004 & 0.501
 \vspace{0.15cm}\\
Reed et. al. (2007b)& 				$0.312^{+0.025}_{-0.022}$   &$0.806^{+0.039}_{-0.030}$& 8.604 & 15.004 & 0.501
 \vspace{0.15cm}\\
 Reed et. al. (2003)&  				$0.328^{+0.025}_{-0.022}$   &$0.789^{+0.037}_{-0.029}$& 8.605 & 15.005 & 0.502
\vspace{0.15cm}\\
Crocce et. al. (2010)& 				$0.318^{+0.024}_{-0.022}$   &$0.758^{+0.036}_{-0.028}$& 8.625 & 15.024 &0.521
\vspace{0.15cm}\\
Maggiore \& Riotto (2010)&		$0.315^{+0.024}_{-0.022}$   &$0.802^{+0.037}_{-0.029}$& 8.638 & 15.038 & 0.535
  \vspace{0.15cm}\\
Benson (2016)& 				$0.307^{+0.023}_{-0.021}$   &$0.841^{+0.039}_{-0.031}$& 8.676 & 15.076 &0.573
\vspace{0.15cm}\\
Despali et. al. (2016)& 			$0.362^{+0.027}_{-0.025}$   &$0.867^{+0.040}_{-0.033}$& 8.798 & 15.198 & 0.695
\vspace{0.15cm}\\
Courtin et. al. (2010)& 		  $0.306^{+0.023}_{-0.021}$   &$0.761^{+0.035}_{-0.028}$& 8.821 & 15.221 & 0.718
 \vspace{0.15cm}\\
Yahagi et. al. (2004)& 				$0.351^{+0.026}_{-0.024}$   &$0.745^{+0.036}_{-0.028}$& 8.842 & 15.242 & 0.739 
 \vspace{0.15cm}\\
 Reed et. al. (2007a)& 				$0.314^{+0.024}_{-0.021}$   &$0.766^{+0.035}_{-0.026}$& 8.884 & 15.284 & 0. 781
 \vspace{0.15cm}\\
 White  (2002)&				 	$0.334^{+0.025}_{-0.023}$   &$0.770^{+0.037}_{-0.029}$& 8.884 & 15.284 & 0.781
 \vspace{0.15cm}\\
 Sheth \& Tormen (1999)&  				$0.335^{+0.025}_{-0.023}$   &$0.797^{+0.038}_{-0.030}$& 8.891 & 15.291 & 0.788
 \vspace{0.15cm}\\
Manera et. al. (2010)& 				$0.311^{+0.024}_{-0.021}$   &$0.765^{+0.036}_{-0.028}$& 8.965 & 15.365 & 0.862
 \vspace{0.15cm}\\
Bagla et. al. (2009)& 				$0.304^{+0.023}_{-0.021}$   &$0.762^{+0.035}_{-0.028}$& 8.982 & 15.382 & 0.879
 \vspace{0.15cm}\\ 
 Press \& Schechter (1974)& 	 	   $0.238^{+0.019}_{-0.017}$   &$0.918^{+0.041}_{-0.032}$& 9.124 & 15.524 & 1.021
 \vspace{0.01cm}\\

\hline\hline
\label{tab:growth2}
\end{tabular}
\end{table*}

\begin{figure}
\includegraphics[width=\linewidth]{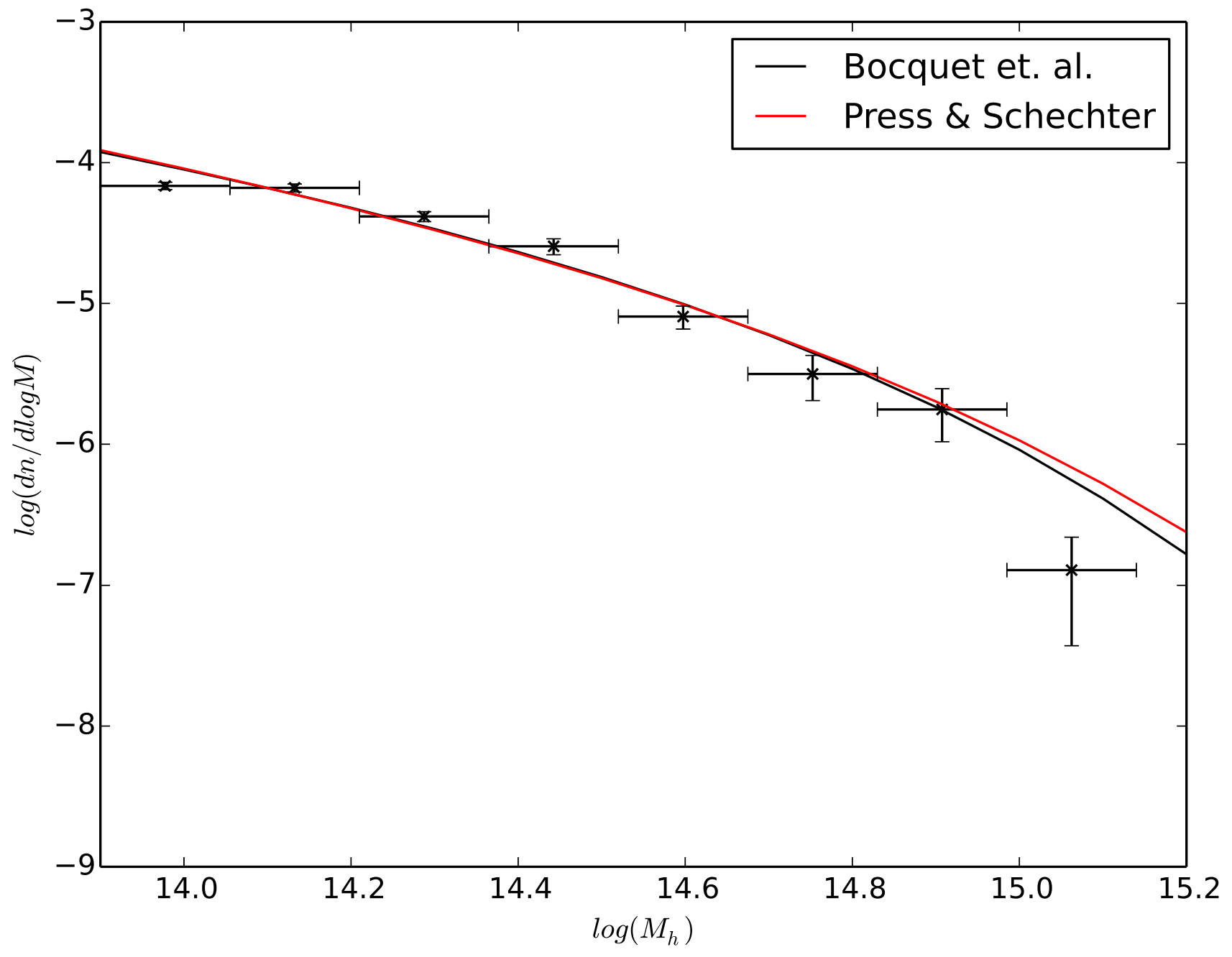}
\caption{Comparison of the Cluster Mass Function data with the theoretical HMF fits of Bocquet et. al. and Press \& Schechter models.}
\label{f1}
\end{figure}

To have a visual appreciation of the results, we present in Figure (\ref{f1}) the comparison of the observed CMF (Table \ref{tab01}) with the most and least successful HMF models, i.e., the \citet{Bocquet16} and the \citet{Press74} models, respectively, utilizing the best fit parameters values provided in Table \ref{tab:growth2}.

\subsection{Comparison of the CMF and Planck-CMB parameters. Is there a $\sigma_{8}$ tension?}

Although the statistical comparison of the observed CMF (Table \ref{tab01}) with the different theoretical HMF models cannot allow us to constrain the range of viable models, it provided a relatively narrow range of the fitted $\Omega_{m,0}$ and $\sigma_8$ cosmological parameters, where 
$0.238\leq \Omega_m \leq 0.362$ and $0.737\leq \sigma_8 \leq 0.918$.
 The weighted (with $\w_i=1/\chi^2_{i, \rm min}$) mean values, over all the HMF models,  are:
\begin{equation}
{\bar \Omega_m} = \frac{\sum{w_i\Omega_{m,i}}}{\sum{w_i}}= 0.313 \pm 0.022
\end{equation}
and
\begin{equation}
{\bar \sigma_{8}} = \frac{\sum{w_i\sigma_{8,i}}}{\sum{w_i}} = 0.798\pm 0.040 \;,
\end{equation}
where the uncertainties represent $1\sigma$ error of the mean values over all HMF models. The corresponding $S_8$-value is $S_8=0.815\pm 0.05$ which is fully consistent, within $\sim 0.3\sigma$, with the Planck-CMB results. Note also that our best fit HMF model (Bocquet et al.) provides almost identical results, ie., $S_8=0.811\pm 0.045$.

Furthermore, we wish to identify those HMF models that are consistent (or inconsistent) with the current Planck-CMB results \citep{Aghanim20}, at a statistically significant level. To achieve that, we compare the best fit results of $\Omega_m$ and $\sigma_8$ parameters, that we obtain for each HMF model (see Table \ref{tab:growth2}), to the Planck-CMB results. More specifically, we compare the results of: (i) {\em Planck} (TT, TE, EE+lowE+lensing) cosmology [Planck], which provides $\Omega_m = 0.315\pm0.007$ and $\sigma_8 = 0.811\pm0.006$; (ii) the {\em Planck} (TT, TE, EE+lowE+lensing) cosmology, combined with BAO [Planck + BAO], which provides $\Omega_m = 0.311\pm0.0056$ and $\sigma_8 = 0.810\pm0.006$; and finally (iii) the {\em Planck} (TT, TE, EE+lowE+lensing) cosmology combined with DES [Planck + DES], which provides $\Omega_m = 0.304\pm0.006$ and $\sigma_8 = 0.8062\pm0.0057$ \citep{Aghanim20}.  
We utilize the index of inconsistency (IOI) method to perform this comparison. The IOI is sensitive to the separation distance of the means, the size of the parameter space, and the orientation of the parameters space (\citet{Lin17}). The IOI is given by
\begin{equation}
IOI\equiv\frac{1}{2}{\bf{\delta^TG\delta}} \;,
\end{equation}
where $\delta = \mu_2 - \mu_1$ is the difference of the means of the parameters provided by the HMF model and by the cosmology, while $\bf{G = (C_1 + C_2)^{-1}}$ is the inverse sum of the covariance matrices of the HMF model parameters and the cosmology. Higher IOI implies higher inconsistency. More specifically, for $IOI<1$ there is no significant inconsistency, for $1<IOI<2.5$ there is a weak inconsistency and for $2.5<IOI<5$ and $IOI>5$  there is a moderate and strong inconsistency, respectively. In Table \ref{tab:growth1} we present the IOI results and as it can be seen, using the Planck, [Planck-BAO] and [Planck-DES] cosmologies there are twenty out of the twenty-three models that show no significant
inconsistency. In the case of \citet{Yahagi_2004} and \citet{Despali_2015} HMFs there is a weak inconsistency, while the \citet{Press74} model shows a strong inconsistency.

Thus out of 23 HMF models there are 20 models that are consistent with both the observed CMF, obtained from \citet{Abdullah20b}, and the joint Planck CMB-BAO and CMB-DES cosmological probes. The inverse-$\chi^2_{min}$ weighted average values of the two cosmological parameters are ${\bar \Omega_{m,0}}=0.312\pm 0.011$ and ${\bar \sigma_8}=0.791\pm0.027$. We also obtain similar results but with a significantly smaller scatter when we use the inverse-IOI weighting scheme. We find ${\bar \Omega_{m,0}} = 0.312 \pm0.003$ and ${\bar \sigma_8} = 0.805 \pm0.007$, corresponding to $S_8=0.821\pm 0.008$.

Our results present no evidence for a $\sigma_8$-tension between the {\em Planck}-CMB and the Cluster Mass Function probe, based on the large majority of the theoretical HMF models, used in this paper.
Rather, the $\sigma_8$-tension between the lower redshift cosmological probes and the CMB could be due to systematic uncertainties of the lower redshift probes and it should probably not be attributed to a weakness of the $\Lambda$CDM model. 
As discussed in \citet{Abdullah20b}, studies that use galaxy clusters to constrain the cosmological parameters $\Omega_m$ and $\sigma_8$, provide conflicting results. Some of these studies are in agreement with the Planck data and some other studies are not. 
These discrepancies could be caused by many reasons among which the method used to calculate the cluster mass, mis-identification of the cluster locations and cluster center, or due to redshift uncertainties.

\begin{table*}%[ht]
\caption[]{IOI results for the mass function data (see Table \ref{tab01}): The
  $1^{st}$ column indicates the mass function model, the
  $2^{nd}$, $3^{rd}$ and the $4^{th}$ column present the IOI results between the models and the {\em Planck},
  Planck+DES and  Planck+BAO cosmologies, respectively.}

\tabcolsep 4.5pt
\vspace{1mm}
\begin{tabular}{cccccc} \hline \hline
 Mass Function Model & {\em Planck} & Planck+DES &  Planck+BAO & \vspace{0.05cm}\\ \hline
Bocquet et. al.(2016)& 				$0.041$             &$0.078$&  $0.045$ \vspace{0.15cm}\\ 
Watson et. al. (2013-FOF)& 		$0.318$             &$0.292$&  $0.307$            \vspace{0.15cm}\\
Jenkins et. al. (2001)& 			$0.196$             &$0.037$&  $0.130$  \vspace{0.15cm}\\
Angulo et. al. (2012)& 				$0.235$             &$0.076$&  $0.172$  \vspace{0.15cm}\\
Bhattacharya et. al. (2011)& 	$0.478$             &$0.402$&  $0.089$  \vspace{0.15cm}\\
Warren et. al. (2006)& 				$0.147$             &$0.017$&  $0.451$  \vspace{0.15cm}\\
Shirasaki et. al. (2021)& 		$0.488$             &$0.685$&  $0.533$  \vspace{0.15cm}\\
Comparat et. al. (2019)& 		  $0.114$             &$0.245$&  $0.140$  \vspace{0.15cm}\\
Tinker et. al.(2008)& 				$0.068$             &$0.010$&  $0.033$ \vspace{0.15cm}\\
Reed et. al. (2007b)& 				$0.001$             &$0.041$&  $0.003$  \vspace{0.15cm}\\
Reed et. al. (2003)&  				$0.158$             &$0.314$&  $0.209$ \vspace{0.15cm}\\
Crocce et. al. (2010)& 				$0.565$             &$0.555$&  $0.505$  \vspace{0.15cm}\\
Maggiore \& Riotto (2010)&		$0.006$             &$0.064$&  $0.015$  \vspace{0.15cm}\\
Benson (2016)& 								$0.249$             &$0.303$&  $0.241$  \vspace{0.15cm}\\
Despali et. al. (2016)& 			$1.419$             &$1.951$&  $1.600$  \vspace{0.15cm}\\
Courtin et. al. (2010)& 		  $0.550$             &$0.423$&  $0.505$  \vspace{0.15cm}\\
Yahagi et. al. (2004)& 				$1.400$             &$1.540$&  $1.504$  \vspace{0.15cm}\\
Reed et. al. (2007a)& 				$0.416$             &$0.385$&  $0.405$  \vspace{0.15cm}\\
Sheth \& Tormen (1999)&  				$0.200$             &$0.423$&  $0.274$  \vspace{0.15cm}\\
White  (2002)&				 				$0.467$             &$0.620$&  $0.521$  \vspace{0.15cm}\\
Manera et. al. (2010)& 				$0.413$             &$0.356$&  $0.392$  \vspace{0.15cm}\\
Bagla et. al. (2009)& 				$0.544$             &$0.400$&  $0.493$  \vspace{0.15cm}\\

Press \& Schechter (1974)& 		$6.100$            &$5.367$& $5.916$ \vspace{0.01cm}\\

\hline\hline
\label{tab:growth1}
\end{tabular}
\end{table*}

\subsection{Robustness of our results: dependence on $h$ and $\Omega_{b} h^{2}$}
Although the cluster abundances as encapsulated in the HMF modelling are mostly sensitive to the $\Omega_m$ and $\sigma_8$ cosmological parameters, there are also other parameters, such as $h$ and $\Omega_bh^2$, that enter in such modelling via the Power-Spectrum of matter fluctuations. 

\begin{figure*}\hspace{0cm}
\centering
\includegraphics[width=\linewidth]{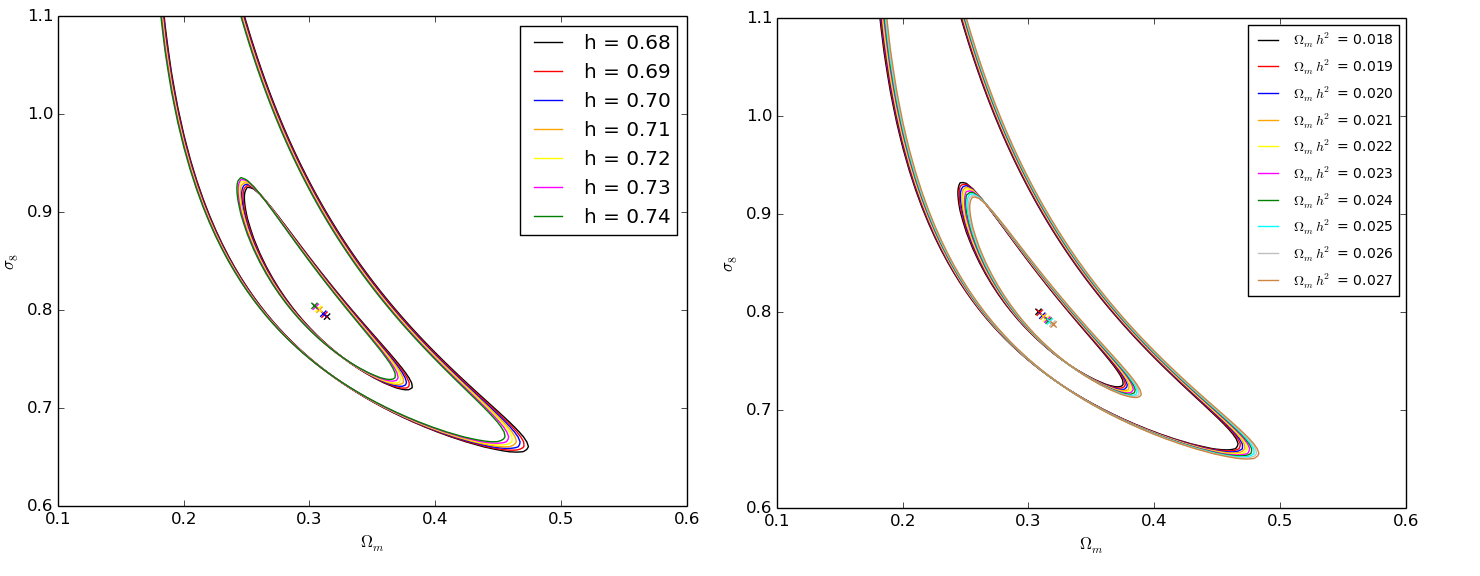}
\caption{Dependence of the cosmological parameters, $\Omega_m$ and $\sigma_8$ derived by fitting the current CMF to the Bocquet et. al. HMF, on the parameters $h$ and $\Omega_bh^2$ that enter in the definition of the matter perturbations power-spectrum. The left plot the $1\sigma$ and $3\sigma$ confidence levels of the two fitted parameters, for different values of $h$. The right plot is corresponding plot, but for different values of $\Omega_bh^2$.  
    }
    \label{f3}
\end{figure*}

In this section we investigate the sensitivity of our results on the variations of $h$ and $\Omega_{b} h^{2}$. We use our best fit model (Bocquet et. al.) and repeat our analysis by allowing $h$ to vary within the range of $0.68\leq h\leq0.74$, in steps of 0.01. We also allow $\Omega_{b} h^{2}$  to vary within the range of $0.018\leq\Omega_bh^2\leq0.027$, in steps of 0.001. The fitted values of the two main cosmological parameters, $\Omega_m$ and $\sigma_8$, show insignificant dependence on the specified range of the $h$ and $\Omega_bh^2$ parameters (see Figure \ref{f3}). This indicates that the effects of varying $h$ and $\Omega_{b} h^{2}$ on the constraints of  $\Omega_m$ ($\sim 4\%$) and $\sigma_8$ ($\sim 2\%$) is minimal.

\subsection{Comparison with other Cluster MF studies}
Utilizing the probe of the cluster mass function, the constraints on the cosmological parameters $\Omega_m$ and $\sigma_8$, has been extensively studied in the literature based on a variety of galaxy surveys in the optical, infrared, X-ray, and mm-wave bands. The basic result is that the inferred (or directly estimated) value of cluster normalization parameter, $S_8$, is systematically lower than that of the CMB temperature fluctuation analyses. Although, due to the relatively large uncertainties of the derived values of $\Omega_m$ and $\sigma_8$ and/or $S_8$, the individual significance is rather low, unless one would combine jointly with other cosmological probes.

\begin{figure}
\includegraphics[width=\linewidth]{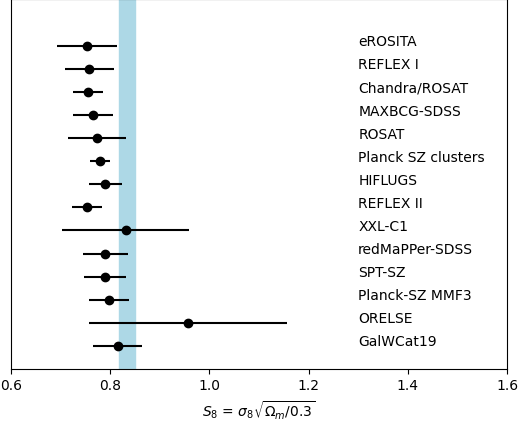}
\caption{Comparison of the derived or inferred $S_8$ values based on the CMF or counts analysis from a number of studies (indicated by the cluster sample analysed) with the {\em Planck} 2018 CMB results (shaded area).}
\label{fmf}
\end{figure}

In Figure (\ref{fmf}) we present a comparison of our results and those of other studies, which are based solely on the cluster mass function or cluster number counts (\citealp{2023MNRAS.522.1601C,Schuecker03,2009ApJ...692.1060V,2010ApJ...708..645R,2015MNRAS.446.2205M,2020A&A...641A...1P, 2017MNRAS.471.1370S,Bohringer17,pac,2019MNRAS.488.4779C,Bocquet19,Zubeldia19,2021MNRAS.502.3942H}), with the CMB {\em Planck} 2018 results (shaded area). 

Note that for the studies where the results of $\Omega_m$ and $\sigma_8$ were presented individually, we transformed them to the pivot $S_8=\sigma_8\sqrt{\Omega_m/0.3}$ relation. Furthermore, in order to have a consistent comparison of the different studies, we ignored results obtained from the joint analyses of the Cluster Mass Function and other cosmological probes.

\section{Conclusions}
\label{sec:Conc}
In this work we investigate the ability of 23 theoretical halo mass function models to represent the observational mass function of galaxy clusters, obtained from  \citet{Abdullah20b}. To this end we apply a standard $\chi^2$ minimization procedure between the models and the observational cluster mass function and constrain the two free cosmological parameters of the models, namely $\Omega_m$ and $\sigma_8$.
Utilizing the best-$\chi^{2}$ and the value of the Akaike information criterion, we find that all the models fit the data at a statistically acceptable level. We find that the \citealp{Bocquet16} halo mass function model returns the smaller $\chi^{2}$ value and therefore it presents the best fit model to the CMF data. We compare all other theoretical HMFs to the reference HMF model \citep{Bocquet16} and find that all models are consistent with it. This fact implies that in order to constrain further among the HMF models we need more observational data of wider cluster-mass and redshift ranges. Since all our models pass the CMF-HMF comparison test at a statistically acceptable level, we determine the average values of ${\bar \Omega_{m},0}$ and ${\bar \sigma_8}$ over all HMF models, weighted by the inverse $\chi^2_{\rm min}$ value. We find that  ${\bar \Omega_{m,0}} = 0.313 \pm0.022$ and ${\bar \sigma_8} = 0.798 \pm0.040$, with corresponding $S_8=0.815\pm 0.05$.

Furthermore in order to identify possible HMF models which are in disagreement with the recent Planck-CMB results, we utilized the IOI method to compare the two fitted free parameters of the models ($\Omega_m, \sigma_8$) with the corresponding parameters obtained from the Planck CMB analysis in addition to the joint Planck+BAO and Planck+DES analyses and find that only three HMF models are inconsistent. These models are the \citet{Yahagi_2004}, \citet{Despali_2015}, and \citet{Press74} HMFs, with the latter being the most inconsistent one. 
Using the 20 models which are consistent with the {\em Planck}-CMB results, we find that the average values of ${\bar \Omega_{m,0}}$ and ${\bar \sigma_8}$, weighted by the inverse-IOI, are ${\bar \Omega_{m,0}} = 0.312 \pm0.003$ and ${\bar \sigma_8} = 0.805 \pm0.007$.

Therefore, we conclude that the current cluster mass function, obtained by \citet{Abdullah20b} at redshift $z\sim 0.1$, provides results that show no $\sigma_8$-tension with the Planck-CMB results within the $\Lambda$CDM cosmological model frame, a result which holds for a large majority of the theoretical halo mass functions used to model the CMF data.

\section*{Data availability}
No new data were generated or analyzed in support of this research.

\bibliographystyle{mnras}
\bibliography{ref1.bib}

\end{document}